\pdfoutput=1

\documentclass[11pt]{article}

\usepackage[final]{acl}
\usepackage{booktabs}

\usepackage{times}
\usepackage{latexsym}

\usepackage[T1]{fontenc}

\usepackage[utf8]{inputenc}

\usepackage{microtype}

\usepackage{inconsolata}

\usepackage{graphicx}

\usepackage{booktabs}
\usepackage{graphicx}
\usepackage{amsmath}
\usepackage{amssymb}
\usepackage{comment}
\usepackage{tcolorbox}
\usepackage{xcolor}
\usepackage{adjustbox}
\usepackage{relsize}

\title{\textsc{MolErr2Fix}: Benchmarking LLM Trustworthiness in Chemistry via Modular Error Detection, Localization, Explanation, and Revision}

\author{
  \textbf{Yuyang Wu\textsuperscript{1}\thanks{Equal contribution}},
  \textbf{Jinhui Ye\textsuperscript{1,2,*}},
  \textbf{Shuhao Zhang\textsuperscript{1,*}},
  \textbf{Lu Dai\textsuperscript{2}},
\\
  \textbf{Yonatan Bisk\textsuperscript{1}\thanks{Corresponding authors}},
  \textbf{Olexandr Isayev\textsuperscript{1,$\dagger$}}
\\
\\
  \textsuperscript{1} Carnegie Mellon University
\\
  \textsuperscript{2}Hong Kong University of Science and Technology
\\\\
\small\texttt{\{yuyangwu, jinhuiy, shuhaozh\}@andrew.cmu.edu, ldaiae@connect.ust.hk,}
\\
\small\texttt{ybisk@cs.cmu.edu, olexandr@cmu.edu}
}

\begin{document}
\maketitle

\begin{abstract}

Large Language Models (LLMs) have shown growing potential in molecular sciences, but they often produce chemically inaccurate descriptions and struggle to recognize or justify potential errors. This raises important concerns about their robustness and reliability in scientific applications. To support more rigorous evaluation of LLMs in chemical reasoning, we present the \textsc{MolErr2Fix} benchmark, designed to assess LLMs on error detection and correction in molecular descriptions.
Unlike existing benchmarks focused on molecule-to-text generation or property prediction, \textsc{MolErr2Fix} emphasizes fine-grained chemical understanding. It tasks LLMs with identifying, localizing, explaining, and revising potential structural and semantic errors in molecular descriptions. Specifically, \textsc{MolErr2Fix} consists of 1,193 fine-grained annotated error instances. Each instance contains quadruple annotations, i.e,. (error type, span location, the explanation, and the correction). 
These tasks are intended to reflect the types of reasoning and verification required in real-world chemical communication.
Evaluations of current state-of-the-art LLMs reveal notable performance gaps, underscoring the need for more robust chemical reasoning capabilities. MolErr2Fix provides a focused benchmark for evaluating such capabilities and aims to support progress toward more reliable and chemically informed language models. All annotations and accompanying evaluation code are publicly available to facilitate future research.


\end{abstract}

\section{Introduction}

\begin{figure}[t]
\includegraphics[width=\columnwidth]{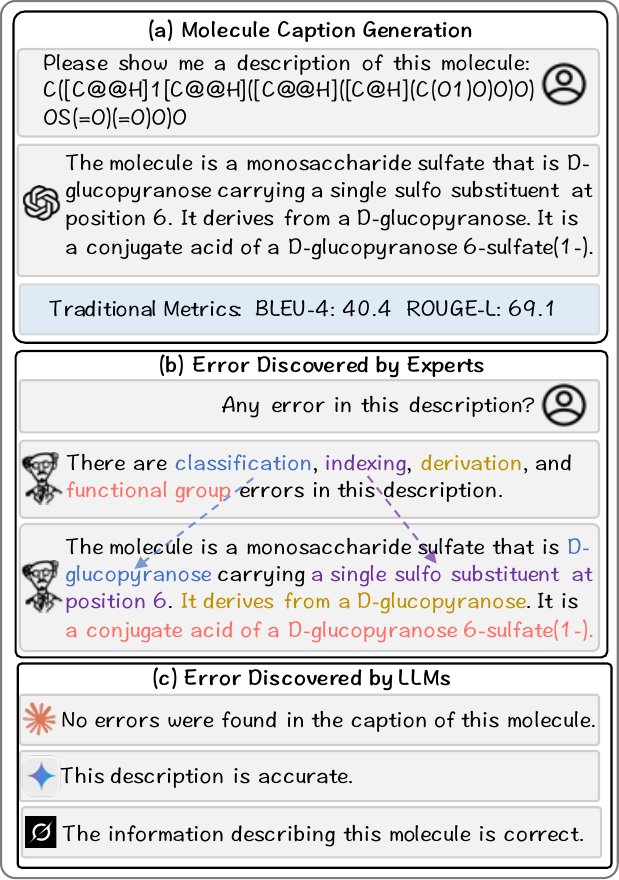}
    \caption{(a) and (b) indicate that the molecular caption generated by LLMs \textbf{exhibits many errors}, even though it has high BLEU and ROUGE scores against the ground truth. (c) indicates LLMs \textbf{fail to detect errors}. }
    \label{fig:figure1}
\end{figure}
Large Language Models (LLMs) have achieved remarkable success in natural language tasks, but recent studies highlight significant limitations when these models are applied to complex chemistry problems \cite{jablonka2022gpt, hatakeyama2023prompt, bran2025chemical}. Even state-of-the-art LLMs struggle with domain-specific knowledge and often produce misleading or incorrect outputs in molecular science applications \cite{guo2023can}. One prominent issue is the hallucination of incorrect details when LLMs describe molecules (“molecular captioning”). In particular, models fine-tuned to translate between chemical structures and text frequently generate factually incorrect or nonsensical information about molecular structures or their properties \cite{lu2024moleculeqa}.
As illustrated in Figure~\ref{fig:figure1}, LLMs often produce fluent and grammatically correct molecular captions, yet misinterpret the underlying SMILES \cite{weininger1988smiles} strings—{for example,} failing to recognize the correct number of atoms or functional groups. Such chemical errors are substantially more critical than surface-level linguistic inaccuracies. More alarmingly, most LLM models consistently fail to recognize their own factual or logical errors in a given translation result \cite{kamoi2024can}.

Previous approaches to evaluate the chemical understanding of models have predominantly relied on indirect assessments through downstream tasks such as \textit{molecular property prediction} or \textit{molecular question answering}~\cite{lu2024moleculeqa}. This reliance on indirect evaluation was, in part, a consequence of the limited generative capabilities of earlier models and a scarcity of sophisticated tools for effectively assessing the quality of generated molecular text: traditional text generation metrics like BLEU~\cite{papineni2002bleu} and ROUGE~\cite{lin2004rouge} prioritized linguistic overlap over restricted outputs, obscuring whether a model genuinely reasons over molecular semantics.
Recently, with the significant advancements in LLM generation, efforts have emerged to directly probe their understanding by having them produce textual descriptions of molecules (e.g., SMILES$\rightarrow$text). Platforms like MolCap-Arena\cite{edwards2024molcap}, for example, facilitate the comparison of different models' descriptive capabilities through ranking methodologies such as pairwise comparisons, but still lack the granularity to provide deep insights into the specific nature of an LLM's misunderstandings of molecular features. They typically do not pinpoint whether a model fails to identify functional groups, miscounts atoms, or makes other critical chemical errors, thus offering limited actionable feedback for targeted model improvement.

In real-world scenarios, LLMs are expected to serve as collaborative chemical assistants—working with scientists, generating hypotheses, and accelerating discovery \cite{bran2025chemical}. Fulfilling this role requires more than fluent molecule descriptions or one‐shot answers; models must also \emph{detect, localize, explain, and revise} their own chemical errors. However, no existing benchmark reveals whether an LLM can self-verify or correct flawed chemical claims—an ability that underpins any trustworthy scientific assistant.

To address this gap, we introduce \textsc{MolErr2Fix}, a structured benchmark that systematically evaluates LLMs’ capabilities in chemical error detecting and reasoning. Given an initial molecular description, LLMs are tasked with: 1) detecting whether errors exist, 2) localizing the erroneous span within the text, 3) explaining the violated chemical principle, and 4) generating a corrected, chemically valid revision.
Unlike existing evaluations focused on fluency or factual recall, \textsc{MolErr2Fix} probes deeper reasoning and diagnostic abilities. It offers fine-grained insights into whether LLMs can leverage domain knowledge to identify and correct errors in molecular descriptions.

By leveraging this comprehensive benchmark, our evaluation reveals critical shortcomings in current models (including GPT-4o and chemistry-specific ChemLLM~\cite{zhang2024chemllm}). While a few models occasionally succeed at detecting errors, they frequently fail to localize, explain, or correct them accurately. These findings underscore that even the most advanced LLMs lack some mechanisms for chemical error discovery and recovery.

Our main contributions can be summarized as follows.
\begin{itemize}
    \item We propose \textsc{MolErr2Fix}, 
     the first benchmark explicitly designed to evaluate LLMs’ ability to detect and recover from chemical errors in molecular descriptions. Unlike prior tasks focused solely on generation or classification, our benchmark directly assesses models’ diagnostic reasoning and reliability in a chemically grounded setting.
     
    \item \textsc{MolErr2Fix} introduces a structured chain-like, four-stage \textit{Error-to-Fix} evaluation pipeline—detection, localization, explanation, and {revision}—offering a fine-grained and progressive framework for assessing LLMs in chemical reasoning.
    \item We provide detailed error analysis and propose evaluation protocols to benchmark and progress toward chemically reliable LLMs.
\end{itemize}


\section{Related Works}
\subsection{Molecule Captioning and Translation}
Early work on bridging molecules with natural language established tasks like molecule captioning and cross-modal retrieval, with \citet{edwards2021text2mol} introducing ChEBI-20 and Text2Mol for molecule-text alignment. Various cross-modal models \citep{ su2022molecular, christofidellis2023unifying, luo2023molfm, liu-etal-2023-molxpt, liu2023molca, zhang2024chemllm, xue2025phyt2v} bridged molecular and natural language via seq2seq tasks (e.g., molecule captioning, de novo generation) and contrastive tasks (e.g., cross-modal retrieval). Generative models like MolT5 \citep{edwards2022translation} and BioT5 \citep{pei2023biot5}, and contrastive models like MoMu \citep{su2022molecular} and MoleculeSTM \citep{liu2023multi}, advanced these tasks. MolReGPT \citep{li2025large} enabled in-context learning with GPT-3.5/4. However, these works focus on generation or retrieval, not chemical correctness.

\subsection{Large Language Models for Chemistry}
Studies on LLMs’ chemical understanding have examined tasks like retrosynthesis planning and functional group identification. \citet{bran2025chemical} integrated GPT-4 with algorithmic search to enable retrosynthesis planning, demonstrating advanced synthetic route analysis. However, \citet{yik2024chatgpt} found LLMs frequently generate incorrect chemical outputs despite fluency. Similarly, \citet{li2024empowering} identified factual errors in MolT5-generated descriptions, undetected by surface-level metrics. MoleculeQA \citep{lu2024moleculeqa}, with over 60,000 questions, showed that specialized models achieve < 50\% accuracy in factual comprehension, underscoring persistent challenges to ensure chemical reliability.

Recent LLMs for chemistry include benchmark studies \citep{white2023assessment, guo2023can, jablonka2024leveraging} evaluating GPT-4 and LLaMA, \citet{guo2023can} noting chemically implausible outputs. Fine-tuning datasets like Mol-Instructions \citep{fang2023mol}, Drugchat \citep{liang2023drugchat}, and MolOpt-Instructions \citep{ye2025drugassist} aim to improve LLMs, with limited gains. SMolInstruct \citep{yu2024llasmol} enables LlaSMol that surpasses state-of-the-art LLMs.  Nevertheless, no prior benchmark directly targets the detection and correction of chemical errors in natural language statements. Our MolErr2Fix fills this gap by challenging LLMs to identify and fix molecular misstatements, thereby exposing the remaining barriers to trustworthy chemical reasoning.



\section{The MolErr2Fix Benchmark}

\begin{figure*}[t]
  \centering
  \includegraphics[width=0.9\textwidth]{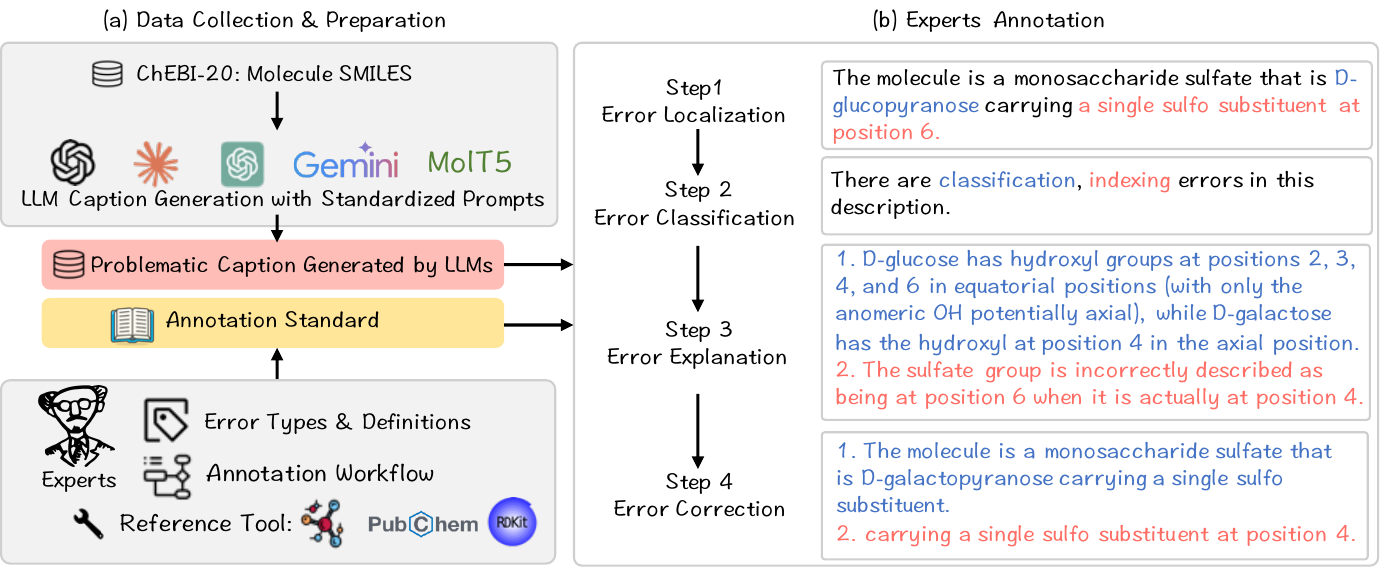}
  \caption{Annotation pipeline of the MolErr2Fix. (a) Problematic molecular candidate captions generation using standardized prompts across multiple LLMs with ChEBI-20 SMILES.  (b) Expert annotation process involves four steps: error localization, classification, explanation, and correction, based on expert-defined taxonomies and reference tools, ensuring chemical accuracy in molecular descriptions.}
  \label{fig:figure2}
\end{figure*}


We introduce MolErr2Fix, a comprehensive molecular benchmark that systematically evaluates Large Language Models (LLMs) by assessing their deep understanding of chemical knowledge through an innovative paradigm of fine-grained error detection, localization, explanation, and correction tasks, which traditionally rely on chemist-level expertise.

\subsection{Task Formulation}
\label{sec:task_formulation}
The MolErr2Fix benchmark evaluates models across four integral tasks, each targeting specific dimensions of chemical reasoning and knowledge. This facilitates a comprehensive assessment of model performance from error identification through to correction. These tasks challenge models in a structured manner, with Figures~\ref{fig:figure2} illustrating details of the four tasks.

\vspace{0.5 em}
\textbf{Error Detection} evaluates whether the model can identify the presence of chemical errors within a given flawed description ($T$) of a molecular structure (SMILES). The model is expected to output a predicted set of error types $\hat{Y}$ that it believes are present, chosen from the six predefined error categories ($\mathcal{C}$). Performance is measured using standard multi-label classification metrics: Precision, defined as $\frac{|\hat{Y} \cap Y|}{|\hat{Y}|}$, and Recall, defined as $\frac{|\hat{Y} \cap Y|}{|Y|}$, where $Y$ is the set of ground-truth error types. The harmonic mean F1 Score is reported to balance these two aspects.

\vspace{0.5 em}
\textbf{Error Localization} measures the model’s ability to locate the specific erroneous text spans in the flawed description ($T$) for a given molecular structure. The model should output a set of predicted error spans $\hat{S} = \{\hat{s}_1, \dots, \hat{s}_n\}$, where each $\hat{s}_j$ denotes a contiguous text segment in $T$ predicted to be erroneous. We evaluate the span-level overlap between these predicted spans and the gold-standard spans $S = \{s_1, \dots, s_m\}$ using Intersection-over-Union (IoU). Specifically, we assess Recall@IoU$\geq\tau$ (the proportion of gold spans $s_i$ for which a predicted span $\hat{s}_j$ achieves an $\text{IoU}(\hat{s}_j, s_i) \geq \tau$, with $\tau$ set at 0.5 and 0.7) and the Average IoU, calculated as $\frac{1}{|S|} \sum_{i=1}^{|S|} \max_j \text{IoU}(\hat{s}_j, s_i)$.

\vspace{0.5 em}
\textbf{Error Explanation} assesses whether the model can explain why a given gold error span $s$ within a flawed description $T$ of a molecule (SMILES) is chemically incorrect. The model is required to generate a concise natural language explanation $\hat{e}$ (typically 1--2 sentences). The quality of these explanations is quantitatively evaluated by comparing them against expert-written references $e$ using GPT-4o as an automatic evaluator for semantic equivalence. This is reported as a Match Rate: $\frac{1}{N} \sum_{i=1}^{N} \mathbb{1}[\hat{e}_i \equiv e_i]$, where $\mathbb{1}[\cdot]$ is an indicator function signifying semantic equivalence as judged by GPT-4o.

\vspace{0.5 em}
\textbf{Error Correction} tests the model’s ability to generate accurate textual corrections for an identified gold error span $s$ in a flawed description $T$ corresponding to a given molecular structure. The model's output is a corrected version $\hat{d}$ that replaces $s$ with a chemically accurate alternative. These corrections are evaluated for textual similarity against expert-validated revisions $d$ using BLEU scores. Furthermore, a Correction Score, based on GPT-4o judging semantic equivalence ($\frac{1}{N} \sum_{i=1}^{N} \mathbb{1}[\hat{d}_i \equiv d_i]$), is used to assess deeper accuracy.

\subsection{Dataset Curation}

The construction of \textsc{MolErr2Fix} follows a two–stage workflow:  
(i)~data collection \& preparation and  
(ii)~expert annotation \& verification, as depicted in Figure \ref{fig:figure2}, with complete prompt templates and decoding parameters provided in the appendix.

\noindent\textbf{Data Collection and Preparation.} 
We begin with the \textsc{ChEBI-20} dataset and select molecules containing fewer than 100 atoms.
This threshold balances molecular complexity and the input–length limits of contemporary sequence-to-sequence generators.

These molecules, represented in Simplified Molecular Input Line Entry System (SMILES) format, were then used to generate captions via LLMs including \textsc{GPT-4o}, \textsc{o3-mini}, \textsc{Claude-Sonnet 3.7}, \textsc{Gemini 1.5}, and the chemistry-specialised \textsc{MolT5-high}, under a zero-shot prompt adapted from MolReGPT.
The resulting free-form descriptions are fluent but frequently contain chemically incorrect spans; we therefore collect them as \textit{Problematic Captions} to be diagnosed in stage (ii).


Meanwhile, to establish a structured annotation scheme, a panel of chemistry experts mapped LLM failure modes onto IUPAC nomenclature and CAS indexing rules, iteratively consulting PubChem until consensus was reached.
The resulting six error types for molecular captioning are: \textit{Functional Group/Substituent Errors} (misidentifying molecular substructures or functional groups), \textit{Classification Errors} (incorrectly categorizing the chemical species), \textit{Derivation Errors} (erroneously linking to chemical precursors or derivatives), \textit{Stereochemistry Errors} (inaccurately assigning stereochemical configurations like R/S or E/Z), \textit{Sequence/Composition Errors} (miscounting molecular components or chain lengths), and \textit{Indexing Errors} (incorrectly locating substituents or features). These error types are not only grounded in established nomenclature practice, but are also designed to comprehensively assess an LLM's depth of understanding across numerical, structural, and semantic aspects of chemical information. This preparatory step laid the groundwork for the subsequent expert annotation stage, where problematic captions would be systematically analyzed and corrected. More detailed definitions and examples of the molecule captioning errors are in the Appendix.

\noindent\textbf{Expert Annotation and Verification.} The second stage involved meticulous annotation and verification by chemistry experts. Guided by predefined annotation criteria and workflows, these experts systematically analyzed the LLM-generated captions against the corresponding molecular structures. For each identified discrepancy, their annotation pipeline comprised of the four key tasks. 

The initial step, \textit{Error Localization}, involved experts precisely pinpointing the specific text spans within the caption that contained an error. Following this, the \textit{Error Classification} task required them to identify the type of chemical error based on the predefined six-category taxonomy: functional group errors, classification errors, derivation errors, stereochemistry inaccuracies, sequence inconsistencies, and indexing errors. Subsequently, for \textit{Error Explanation}, the experts articulated a concise and chemically precise explanation for why the identified span was incorrect, referencing relevant chemical principles or structural features. The final task, \textit{Error Correction}, involved providing a corrected version of the erroneous text span or, if necessary, the entire description, ensuring chemical accuracy and consistency with the reference molecule. 

To ensure the validity of their assessments and corrections, the experts consistently utilized a suite of standard reference tools, including molecular visualization software (e.g., RDKit), authoritative chemical databases (such as PubChem), and relevant research articles. Furthermore, a rigorous verification phase, incorporating a cross-check and recheck protocol among at least two experts for each instance, was implemented to maintain high inter-expert agreement and overall annotation quality. This expert-driven annotation and verification phase was crucial in transforming the raw, potentially flawed LLM outputs into a high-quality benchmark dataset. At the end, each instance includes the molecule, the erroneous description, and detailed, verified annotations regarding the error's type, location, explanation, and correction.

\subsection{Dataset Statistics}
\begin{table}[!t]
\centering

\resizebox{\columnwidth}{!}{%
\begin{tabular}{lc}
\toprule
\textbf{Statistics} & \textbf{Number} \\
\midrule
Total Molecules & 525 \\
Total Annotated Errors & 1193 \\
Error Types & 6 \\
Models Evaluated & 5 \\
\midrule
\multicolumn{2}{l}{\textit{Error Type Distribution}} \\
\midrule
Functional Group/Substituent Errors & 318 (26.6\%) \\
Classification Errors & 303 (25.4\%) \\
Derivation Errors & 220 (18.4\%) \\
Sequence/Composition Errors & 152 (12.7\%) \\
Stereochemistry Errors & 111 (9.3\%) \\
Indexing Errors & 89 (7.5\%) \\
\midrule
\multicolumn{2}{l}{\textit{Error Frequency}} \\
\midrule
Peak Errors per Molecule & 3 ($\sim$225 instances) \\
Average Errors per Molecule & 2.27 \\
\midrule
\multicolumn{2}{l}{\textit{Textual Component Length (characters)}} \\
\midrule
Average Description Length & 652.2 \\
Average Error Segment Length & 419.8 \\
Average Explanation Length & 167.6 \\
Average Correction Length & 211.0 \\
\bottomrule
\end{tabular}%
}
\caption{Key Statistics of the MolErr2Fix Benchmark}
\label{tab:molerr2fix_stats}
\vspace{-0.3cm}
\end{table}

The MolErr2Fix benchmark is constructed from 525 unique molecules, leading to 1,193 meticulously annotated errors, categorized into six distinct types, as shown in Table ~\ref{tab:molerr2fix_stats}. The distribution of these errors highlights specific challenges for Large Language Models (LLMs) in the chemical domain. \textit{Functional Group/Substituent Errors} are the most prevalent, accounting for $26.6\%$ of all identified issues, closely followed by \textit{Classification Errors} at $25.4\%$. This indicates that LLMs often struggle with accurately identifying and describing fundamental chemical moieties and correctly categorizing molecules. Other error types include \textit{Derivation Errors} ($18.4\%$), \textit{Sequence/Composition Errors} ($12.7\%$), \textit{Stereochemistry Errors} ($9.3\%$), and \textit{Indexing Errors} ($7.5\%$). The textual characteristics of the benchmark components also provide insight: average description lengths are $652.2$ characters, while the average length of an identified erroneous segment is $419.8$ characters. This considerable length for error segments implies that errors are often complex and deeply embedded within the surrounding text. Expert-provided reasoning explanations average $167.6$ characters and corrections average $211.0$ characters, underscoring the detailed nature of the annotations and the effort required for rectification, reflecting the diverse complexity in error justification and correction of our benchmark.

\begin{table*}[h!]
\centering
\smaller[+1]
\setlength{\tabcolsep}{2pt}
\renewcommand{\arraystretch}{0.9}

\begin{tabular}{l ccc ccc cc cc}
\toprule
\textbf{Models} & \multicolumn{3}{c}{\textbf{ErrorDetection}} & \multicolumn{3}{c}{\textbf{ErrorLocalization}} & \multicolumn{2}{c}{\textbf{ErrorExplanation}} & \multicolumn{2}{c}{\textbf{ErrorRevision}} \\
\cmidrule(r){2-4} \cmidrule(r){5-7} \cmidrule(r){8-9} \cmidrule(r){10-11}
& Recall & Precision & F1 & IoU$_{\geq 0.5}$ & IoU$_{\geq 0.7}$ & IoU$_{avg}$ & BLEU & GPTscore & BLEU & GPTscore \\
\midrule

\multicolumn{11}{c}{\textbf{1) LLM Models}} \\

gpt-4  
  &  9.7 & 48.3 & 16.1  
  & 21.8 & 15.3 & 15.7  
  & 1.4 & 12.8  
  & 9.6 & 3.5 \\

gpt-4o  
  & 52.3 & 49.8 & 51.0  
  & 35.4 & 20.9 & 23.3  
  & 1.6 & 13.7  
  & 4.7 & 1.7 \\

qwenvl-72B  
  & 45.6 & 47.4 & 46.5  
  & 47.2 & 29.1 & 31.7  
  & 0.8 &  3.2  
  & 1.4 &  0.2 \\
  
gemini-2.0-flash  
  & 36.2 & 54.4 & 43.5  
  & 27.9 & 16.1 & 18.2  
  & 1.1 & 10.0  
  & 1.5 & \textbf{12.5} \\
deepseek-chat  
  & 35.0 & 58.3 & 43.8  
  & 50.2 & 34.1 & 35.8  
  & 1.2 & 15.2  
  & 0.9 & 4.5 \\

Claude-3.7-sonnet
  & 51.9 & 53.3 & 52.6
  & 31.6 & 12.8 & 16.1
  & \textbf{1.9} & 20.9
  & 0.5 & 7.9\\

Grok-3
  & \textbf{76.8} & 58.7 & \textbf{66.6}
  & 41.1 & 25.6 & 26.7
  & \textbf{1.9} & 21.6
  & 0.4 & 7.4\\

\hline
\multicolumn{11}{c}{\textbf{2) LLM Reasoner Models}} \\

o3-mini  
  & 29.8 & 56.8 & 39.1  
  & 29.0 & 14.5 & 17.4  
  & 0.8 & 41.1  
  & 1.8 & 2.8 \\

o4-mini  
  & 44.8 & 53.9 & 53.5  
  & 24.1 & 12.7 & 14.9  
  & 0.4 & \textbf{43.1}  
  & 1.6 & 3.6 \\

deepseek-reasoner

  & 52.8 & 54.3 & 48.9  
  & 31.2 & 17.3 & 20.1  
  & 0.4 & 18.4  
  & 1.8      & 2.0  \\

\hline
\multicolumn{11}{c}{\textbf{3) Domain-specific LLM Models}} \\
LlaSMol  
  & 34.2 & 3.4 & 6.2 
  & 0.0 & 0.0 & 0.0
  & 0.0 & 0.0 
  & 0.0 & 0.0 \\
  
ChemLLM-chat  
  & 40.3 & 4.2 & 7.5  
  & 0.0 & 0.0 & 0.0 
  & 0.0 & 0.0 
  & 0.0 & 0.0 \\
ChemLLM-chat-DPO  
  & 100.0 & 4.2 & 8.0  
  & 0.0 & 0.0 & 0.0
  & 0.0 & 0.0 
  & 0.0 & 0.0 \\


\hline
\multicolumn{11}{c}{\textbf{4) LLM Models + 5-shot Prompt}} \\
gpt-4  
  & 25.7 & 51.5 & 34.3  
  & 2.7 & 2.5 & 2.5  
  & 1.6 & 8.8  
  & \textbf{11.1} & 4.5 \\

gpt-4o  
  & {50.7} & 53.9 & 52.3  
  & \textbf{60.1} &\textbf{47.4} & \textbf{49.1}  
  & 1.6 & 12.8  
  & 5.7 & 2.1 \\

o3-mini  
  & 36.5 & \textbf{60.5} & 45.6  
  & 38.5 & 25.3 & 27.4  
  & 0.9 & 20.0  
  & 2.7 & 5.0 \\

o4-mini  
  & 50.3 & 52.4 & 51.3  
  & 41.1 & 28.6 & 30.6  
  & 0.5 & 24.3  
  & 2.5 & 4.6 \\

qwen-plus  
  & 29.0 & 46.4 & 35.7  
  & 39.7 & 29.7 & 31.1  
  & 1.3 & 5.1  
  & 3.2 & 4.8 \\
  
gemini-2.0-flash  
  & 40.0 & 47.1 & 43.2  
  & 55.1 & 45.7 & 47.1  
  & 1.7 & 9.9  
  & 6.1 & 1.8 \\


\bottomrule

\end{tabular}
\caption{Evaluation results of different LLMs on Error Detection, Localization, Explanation, and Revision tasks in the \textsc{MolErr2Fix} benchmark.}
\label{tab:mainresults}
\end{table*}

\vspace{ -0.5 em}
\section{Experiments}
\vspace{ -0.5 em}
The main purpose of baseline experiments is to investigate LLMs' performance in detecting, localizing, explaining, and revising errors in molecular descriptions. The evaluation encompasses four tasks: Error Detection, Error Localization, Error Explanation, and Error Revision.

\subsection{Baselines}

We categorize the baseline models into three groups based on their architectural design and adaptation strategies:

\vspace{0.5 em}
\textbf{General-purpose LLMs}: This group includes \texttt{gpt-4} \cite{achiam2023gpt}, \texttt{gpt-4o} \cite{hurst2024gpt}, \texttt{qwen1-72B} \cite{bai2023qwen}, \texttt{gemini-2.0-flash} \cite{team2023gemini}, \texttt{deepseek-chat} \cite{liu2024deepseek}, \texttt{Claude-3.7-sonnet} \cite{anthropic2024claude}, and \texttt{Grok-3}. These models, developed for broad natural language understanding tasks, serve as a foundation for assessing general error-handling abilities in molecular contexts.

\vspace{0.5 em}
\textbf{Reasoning-enhanced LLMs}: Represented by \texttt{o3-mini}, \texttt{o4-mini}, and \texttt{deepseek-reasoner} \cite{guo2025deepseek}, these models are either designed or fine-tuned to excel in reasoning tasks. They serves to test the hypothesis that enhanced logical inference capabilities improve performance on chemically intricate error-handling tasks.\\

\textbf{Domain-specific LLMs}: This category features models that have been specifically adapted for the chemical domain. We include \texttt{ChemLLM-chat} \texttt{ChemLLM-chat-DPO} \cite{zhang2024chemllm} and \texttt{LlaSMol} \cite{yu2024llasmol} to allow for an assessment of the influence brought by specialized domain knowledge in addressing molecular errors.\\
\textbf{Few-shot learning variants}: We evaluate a subset of models—-\texttt{gpt-4}, \texttt{gpt-4o}, \texttt{gemini-2.0-flash}, \texttt{o3-mini}, \texttt{o4-mini}, and \texttt{qwen-plus}—in a few-shot setting, providing five task-specific examples for each. This category aims to examine the impact of in-context learning on task performance.

These categories enable a comprehensive comparison across model types and adaptation methods. We evaluate the four tasks of the \textsc{MolErr2Fix} benchmark as defined in Section 3.1. For \textit{Error Detection}, we use Precision, Recall, and F1 Score. For \textit{Error Localization}, we report Recall@IoU$\geq0.5$, Recall@IoU$\geq0.7$, and Average IoU. For \textit{Error Explanation} and \textit{Error Revision}, we apply BLEU and GPTScore to measure textual and semantic accuracy, respectively. All models are evaluated in a zero-shot setting, except for the few-shot learning variants. Details about the prompting strategies are provided in the Appendix.

\subsection{Main Results}

This section presents a comprehensive comparison of various LLMs on the \textsc{MolErr2Fix} benchmark, with detailed results presented in Table \ref{tab:mainresults}. Our key findings are summarized as follows:

\noindent\textbf{Performance across tasks}. Error Detection emerged as the most tractable task, where \texttt{Grok-3} achieved the highest F1 Score of 66.6\%, followed by \texttt{Claude-3.7-sonnet} (52.6\%). In contrast, Error Revision proved to be the most challenging task, with no model surpassing a BLEU score of 11.1\% (\texttt{gpt-4}, 5-shot). Error Localization and Error Explanation exhibited intermediate difficulty, with peak performances of 60.1\% Recall@IoU$\geq 0.5$ (\texttt{gpt-4o}, 5-shot) for the former and a 43.1\% GPT Score (\texttt{o4-mini}) for the latter, respectively. This observed hierarchy suggests an increasing level of difficulty for tasks that demand more profound chemical reasoning.

\noindent\textbf{Limitations of domain-specific models}.
We assess three domain-specific 7B models—\texttt{LlaSMol}, \texttt{ChemLLM-chat}, and \texttt{ChemLLM-chat-DPO}—on our benchmark. These models are fine-tuned for narrow tasks like molecular captioning, and lack the dialogue or reasoning abilities required for multi-step instructions. As a result, they perform poorly across all tasks, especially in localization, explanation, and revision. The results highlight that domain alignment alone is insufficient without instruction-following capabilities.

\noindent\textbf{Impact of Reasoning-enhanced}. 
Reasoning-enhanced models, such as \texttt{o4-mini} (F1: 53.5\% in Error Detection) and \texttt{deepseek-reasoner}, consistently outperformed general-purpose LLMs like \texttt{gpt-4} (F1: 16.1\%). This indicates that specialized training for logical inference significantly enhances error-handling capabilities within the molecular domain.

\noindent\textbf{Impact of few-shot learning}. Few-shot learning variants, notably \texttt{gpt-4o} (5-shot), demonstrated marked improvements, particularly in Error Localization, achieving IoU$\geq 0.5$ of 60.1\% compared to 35.4\% in a zero-shot setting. However, the performance gains were limited in Error Explanation and Error Revision, suggesting that in-context learning alone may be insufficient for these more complex generative tasks.

\begin{figure}[t]
  \centering
  \includegraphics[width=0.48\textwidth]{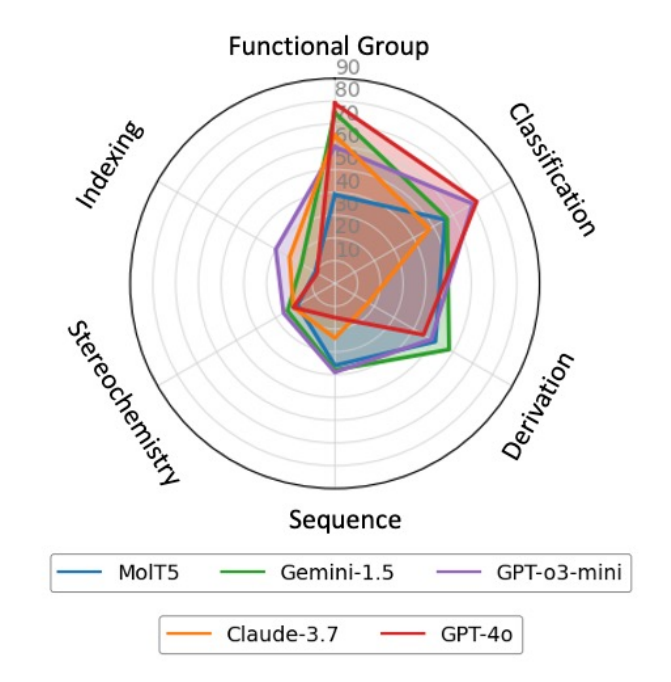}
  \vspace{-0.6 em}
  \caption{Error distribution of six chemical error types in the outputs of five advanced LLMs.}
  \label{Fig:Radar}
\end{figure}
\vspace{-0.5 em}
\subsection{In-depth Analysis of Failure Modes}

While the main results provide a quantitative overview of model performance, a more granular investigation is crucial for understanding the specific weaknesses and challenges in molecular captioning. This section therefore delves into the predominant failure modes, the models' capabilities in error discovery and correction, and an exploration of potential underlying causes for the observed limitations.

\begin{figure}[t]
  \centering
  \includegraphics[width=0.45\textwidth]{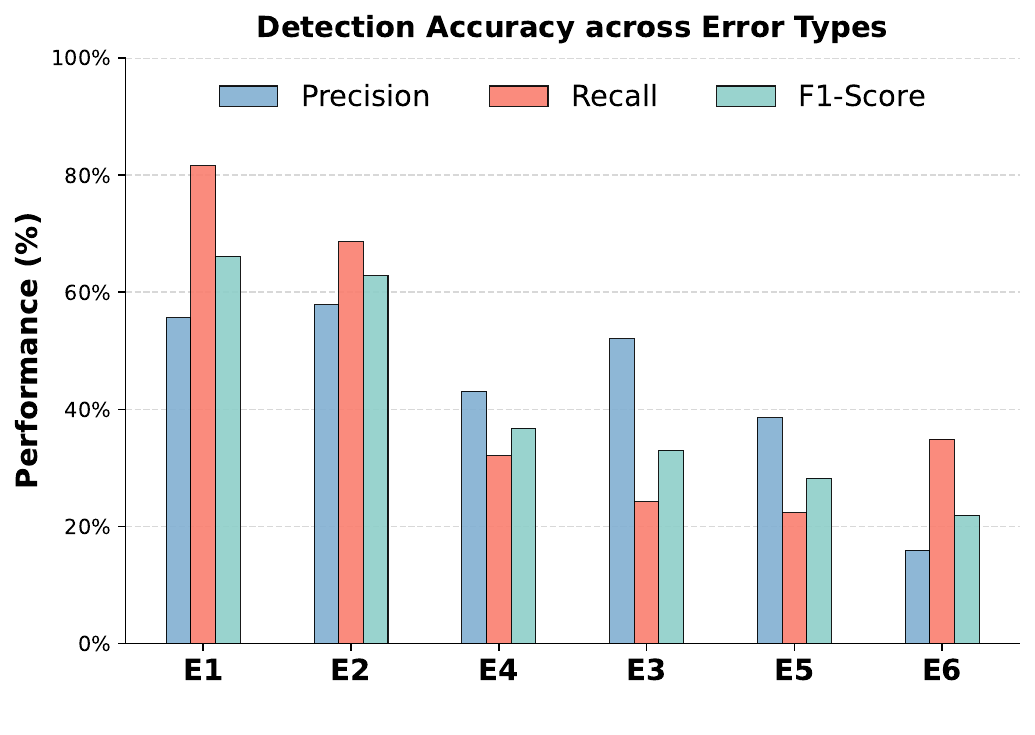}
  \caption{Error detection performance of \texttt{GPT-4o} across six chemical error types in the \textsc{MolErr2Fix} benchmark.}
  \label{Fig:Detection}
\end{figure}
\vspace{-0.5 em}

\subsubsection{Identifying Dominant Failure Modes}

Figure~\ref{Fig:Radar} exposes three systemic weaknesses shared by all five models—Functional Group, Derivation, and Classification errors.  Functional‐group errors are the most pervasive, with GPT-4o and Claude mislabeling local substructures in \(\approx\!75\%\) of cases, underscoring the difficulty of binding graph-level patterns (e.g., sulfonate vs.\ sulfonamide) to precise lexemes.  Derivation errors, frequently exceeding 70\% on Gemini and GPT-4o, reveal an equally stark gap in reasoning over reaction or conjugation semantics: models rarely grasp that \textit{methyl salicylate} is an ester of \textit{salicylic acid} or that deprotonation conserves heavy-atom topology.  Even MolT5—which is pre-trained on chemistry corpora—shows a non-trivial 24 \% misclassification rate, indicating persistent difficulty coupling global topology with formal chemical taxonomies.

\subsubsection{Limits of Error Discovery} 
\begin{figure}[t]
  \centering
  \includegraphics[width=0.35\textwidth]{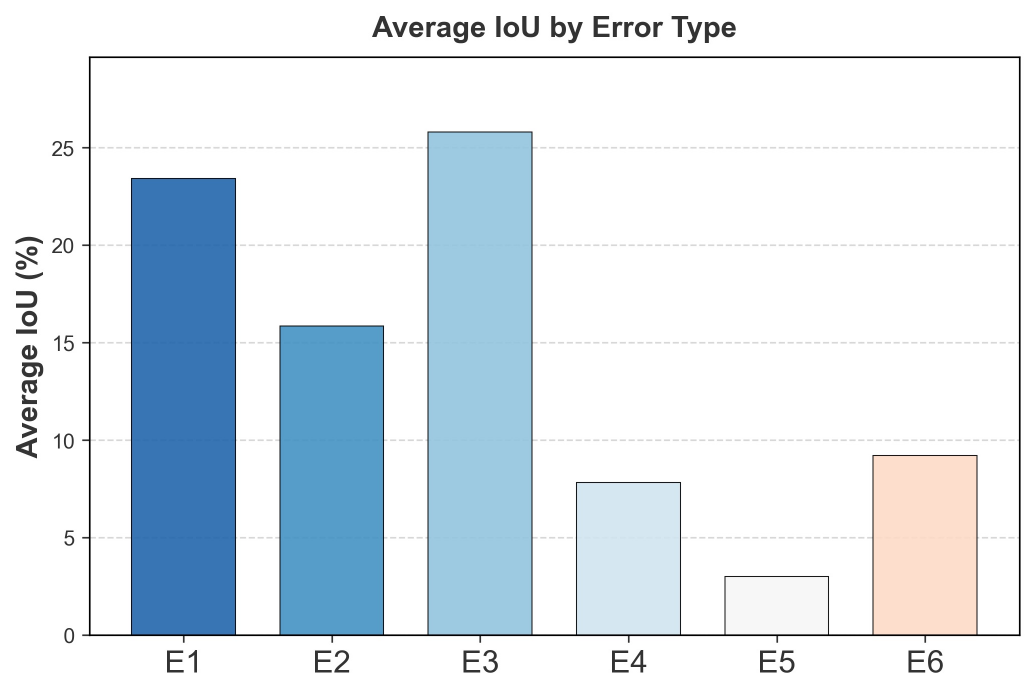}
  \caption{Localization performance of \texttt{GPT-4o} across six chemical error types in the \textsc{MolErr2Fix} benchmark.}
  \label{Fig:Localization}
\end{figure}

The MolErr2Fix results in Figures~\ref{Fig:Detection}–\ref{Fig:Localization} confirm that LLMs detect linguistically salient flaws but falter when numerical or spatial reasoning is required.  Sequence errors (E5) are identified with a poor \( \text{F}_{1}=12.7\%\) and localized with \(\text{IoU}=0.35\%\); Indexing errors (E6) performs even worse (\( \text{F}_{1}=7.9\%\), \(\text{IoU}=6.1\%\)).  These patterns highlight reliance on surface cues: Training data emphasize stylistic fluency, not atom counts or positional descriptors, leaving LLMs blind to subtle violations of IUPAC nomenclature or ring-index conventions.

\subsubsection{Error Recovery Remains Elusive}
\begin{figure}[t]
  \centering
  \includegraphics[width=0.35\textwidth]{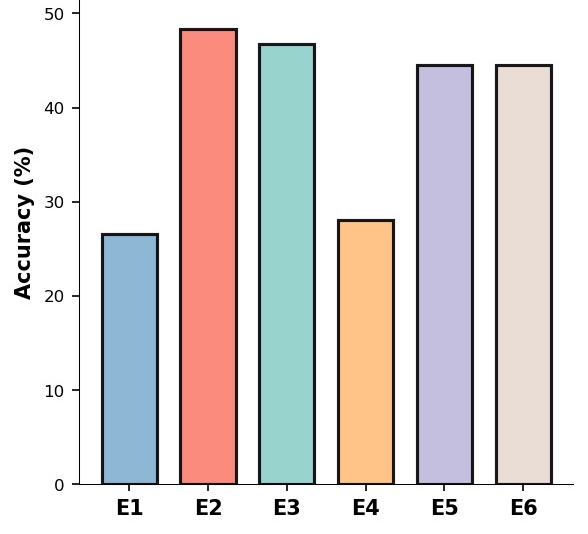}
  \caption{Ability of \texttt{o4-mini} to generate accurate explanations for different categories of chemical errors.}
  \vspace{0.5 em}
  \label{Fig:Explanation}
\end{figure}

Figure~\ref{Fig:Explanation} benchmarks GPT-o4-mini—the best performer in our study—on explanation and revision.  While semantic parity scores (GPT Score \(\approx 43\%\)) suggest it can paraphrase expert rationales, BLEU scores below 2 reveal brittle alignment with gold explanations.  Revision is weaker still: BLEU\(<2\) and GPT Score\(<4\%\) for both E5 and E6 underscore that correcting numeric chains or ortho/meta/para indices remains largely out of reach.

\subsubsection{Exploring Potential Underlying Causes}
While the exact reasons for the low performance of LLMs in molecule captioning remain uncertain, we propose several possible explanations from a chemical perspective:

Certain essential domain knowledge, such as the indexing rules of atoms in fused-ring systems, is rarely detailed even in textbooks, let alone in the corpora typically used to train LLMs. This scarcity of specialized information makes it challenging to identify and address deficiencies in indexing tasks.

Stereochemistry information in SMILES representations is not explicitly indicated by specific symbols; instead, chiral configurations must be inferred from the relative order of fragment representations around an '@' symbol. Such inferential demands pose significant challenges for non-reasoning-based models in correctly distinguishing chiral structures.

Historically, many chemical species and fragments have multiple valid names, creating inherent ambiguities. Such ambiguities may increase the complexity for LLMs to consistently select appropriate terminology.\\

\vspace{-0.5 em}
\vspace{-1 em}

\section{Conclusion and Future Work}
MolErr2Fix introduces a four-stage audit—detection, localization, explanation, and correction—that forces LLMs to reason chemically rather than linguistically. Our evaluation reveals shortcomings in existing models: even state-of-the-art systems spot obvious flaws but falter when asked to locate, justify, or repair them, exposing a gap between fluent language generation and dependable chemical reasoning. We advocate (1) chemistry-centric pretraining architectures, (2) self-reflection loops for iterative debugging, and (3) an expanded benchmark covering richer chemistries and error modes. MolErr2Fix thus maps today’s limits and charts a path toward scientifically trustworthy LLMs.
\section{Limitations}
This study on language models' chemical error detection has several limitations: the dataset only includes molecules under 100 atoms; six predefined error categories don't capture all possible errors; and expert annotations may contain subjective biases. The evaluation is limited to certain model types, uses only a few-shot examples, and relies on metrics that may not fully capture chemical accuracy. Future research should expand molecular complexity coverage, error types, model diversity, and develop chemistry-specific evaluation metrics.
\section{Acknowledgements}

This work was made possible by the Office of Naval Research (ONR) through support provided by the Energetic Materials Program (MURI grant number N00014-21-1-2476). Computation for this work was performed using the high-performance computing clusters Expanse, at San Diego Supercomputer Center (SDSC), and Delta, at National Center for Supercomputing Applications (NCSA), through allocation CHE200122 from the Advanced Cyberinfrastructure Coordination Ecosystem: Services \& Support (ACCESS) program, which is supported by National Science Foundation Grants Numbers 2138259, 2138286, 2138307, 2137603, and 2138296. This research is also a part of the Frontera computing project at the Texas Advanced Computing Center. Frontera is made possible by the National Science Foundation award OAC1818253.
\bibliography{custom}

\begin{thebibliography}{36}
\providecommand{\natexlab}[1]{#1}

\bibitem[{Achiam et~al.(2023)Achiam, Adler, Agarwal, Ahmad, Akkaya, Aleman,
  Almeida, Altenschmidt, Altman, Anadkat et~al.}]{achiam2023gpt}
Josh Achiam, Steven Adler, Sandhini Agarwal, Lama Ahmad, Ilge Akkaya,
  Florencia~Leoni Aleman, Diogo Almeida, Janko Altenschmidt, Sam Altman,
  Shyamal Anadkat, et~al. 2023.
\newblock Gpt-4 technical report.
\newblock \emph{arXiv preprint arXiv:2303.08774}.

\bibitem[{Anthropic(2024)}]{anthropic2024claude}
AI~Anthropic. 2024.
\newblock The claude 3 model family: Opus, sonnet, haiku.
\newblock \emph{Claude-3 Model Card}, 1:1.

\bibitem[{Bai et~al.(2023)Bai, Bai, Chu, Cui, Dang, Deng, Fan, Ge, Han, Huang
  et~al.}]{bai2023qwen}
Jinze Bai, Shuai Bai, Yunfei Chu, Zeyu Cui, Kai Dang, Xiaodong Deng, Yang Fan,
  Wenbin Ge, Yu~Han, Fei Huang, et~al. 2023.
\newblock Qwen technical report.
\newblock \emph{arXiv preprint arXiv:2309.16609}.

\bibitem[{Bran et~al.(2025)Bran, Neukomm, Armstrong, Jon{\v{c}}ev, and
  Schwaller}]{bran2025chemical}
Andres~M Bran, Theo~A Neukomm, Daniel~P Armstrong, Zlatko Jon{\v{c}}ev, and
  Philippe Schwaller. 2025.
\newblock Chemical reasoning in llms unlocks steerable synthesis planning and
  reaction mechanism elucidation.
\newblock \emph{arXiv preprint arXiv:2503.08537}.

\bibitem[{Christofidellis et~al.(2023)Christofidellis, Giannone, Born, Winther,
  Laino, and Manica}]{christofidellis2023unifying}
Dimitrios Christofidellis, Giorgio Giannone, Jannis Born, Ole Winther, Teodoro
  Laino, and Matteo Manica. 2023.
\newblock Unifying molecular and textual representations via multi-task
  language modelling.
\newblock In \emph{International Conference on Machine Learning}, pages
  6140--6157. PMLR.

\bibitem[{Edwards et~al.(2022)Edwards, Lai, Ros, Honke, Cho, and
  Ji}]{edwards2022translation}
Carl Edwards, Tuan Lai, Kevin Ros, Garrett Honke, Kyunghyun Cho, and Heng Ji.
  2022.
\newblock Translation between molecules and natural language.
\newblock \emph{arXiv preprint arXiv:2204.11817}.

\bibitem[{Edwards et~al.(2024)Edwards, Lu, Hajiramezanali, Biancalani, Ji, and
  Scalia}]{edwards2024molcap}
Carl Edwards, Ziqing Lu, Ehsan Hajiramezanali, Tommaso Biancalani, Heng Ji, and
  Gabriele Scalia. 2024.
\newblock Molcap-arena: A comprehensive captioning benchmark on
  language-enhanced molecular property prediction.
\newblock \emph{arXiv preprint arXiv:2411.00737}.

\bibitem[{Edwards et~al.(2021)Edwards, Zhai, and Ji}]{edwards2021text2mol}
Carl Edwards, ChengXiang Zhai, and Heng Ji. 2021.
\newblock Text2mol: Cross-modal molecule retrieval with natural language
  queries.
\newblock In \emph{Proceedings of the 2021 Conference on Empirical Methods in
  Natural Language Processing}, pages 595--607.

\bibitem[{Fang et~al.(2023)Fang, Liang, Zhang, Liu, Huang, Chen, Fan, and
  Chen}]{fang2023mol}
Yin Fang, Xiaozhuan Liang, Ningyu Zhang, Kangwei Liu, Rui Huang, Zhuo Chen,
  Xiaohui Fan, and Huajun Chen. 2023.
\newblock Mol-instructions: A large-scale biomolecular instruction dataset for
  large language models.
\newblock \emph{arXiv preprint arXiv:2306.08018}.

\bibitem[{Guo et~al.(2025)Guo, Yang, Zhang, Song, Zhang, Xu, Zhu, Ma, Wang, Bi
  et~al.}]{guo2025deepseek}
Daya Guo, Dejian Yang, Haowei Zhang, Junxiao Song, Ruoyu Zhang, Runxin Xu,
  Qihao Zhu, Shirong Ma, Peiyi Wang, Xiao Bi, et~al. 2025.
\newblock Deepseek-r1: Incentivizing reasoning capability in llms via
  reinforcement learning.
\newblock \emph{arXiv preprint arXiv:2501.12948}.

\bibitem[{Guo et~al.(2023)Guo, Nan, Liang, Guo, Chawla, Wiest, Zhang
  et~al.}]{guo2023can}
Taicheng Guo, Bozhao Nan, Zhenwen Liang, Zhichun Guo, Nitesh Chawla, Olaf
  Wiest, Xiangliang Zhang, et~al. 2023.
\newblock What can large language models do in chemistry? a comprehensive
  benchmark on eight tasks.
\newblock \emph{Advances in Neural Information Processing Systems},
  36:59662--59688.

\bibitem[{Hatakeyama-Sato et~al.(2023)Hatakeyama-Sato, Yamane, Igarashi, Nabae,
  and Hayakawa}]{hatakeyama2023prompt}
Kan Hatakeyama-Sato, Naoki Yamane, Yasuhiko Igarashi, Yuta Nabae, and Teruaki
  Hayakawa. 2023.
\newblock Prompt engineering of gpt-4 for chemical research: what can/cannot be
  done?
\newblock \emph{Science and Technology of Advanced Materials: Methods},
  3(1):2260300.

\bibitem[{Hurst et~al.(2024)Hurst, Lerer, Goucher, Perelman, Ramesh, Clark,
  Ostrow, Welihinda, Hayes, Radford et~al.}]{hurst2024gpt}
Aaron Hurst, Adam Lerer, Adam~P Goucher, Adam Perelman, Aditya Ramesh, Aidan
  Clark, AJ~Ostrow, Akila Welihinda, Alan Hayes, Alec Radford, et~al. 2024.
\newblock Gpt-4o system card.
\newblock \emph{arXiv preprint arXiv:2410.21276}.

\bibitem[{Jablonka et~al.(2024)Jablonka, Schwaller, Ortega-Guerrero, and
  Smit}]{jablonka2024leveraging}
Kevin~Maik Jablonka, Philippe Schwaller, Andres Ortega-Guerrero, and Berend
  Smit. 2024.
\newblock Leveraging large language models for predictive chemistry.
\newblock \emph{Nature Machine Intelligence}, 6(2):161--169.

\bibitem[{Jablonka et~al.(2022)Jablonka, Schwaller, and Smit}]{jablonka2022gpt}
Kevin~Maik Jablonka, Philippe Schwaller, and Berend Smit. 2022.
\newblock Is gpt-3 all you need for machine learning for chemistry?
\newblock In \emph{AI for Accelerated Materials Design NeurIPS 2022 Workshop}.

\bibitem[{Kamoi et~al.(2024)Kamoi, Zhang, Zhang, Han, and Zhang}]{kamoi2024can}
Ryo Kamoi, Yusen Zhang, Nan Zhang, Jiawei Han, and Rui Zhang. 2024.
\newblock When can llms actually correct their own mistakes? a critical survey
  of self-correction of llms.
\newblock \emph{Transactions of the Association for Computational Linguistics},
  12:1417--1440.

\bibitem[{Li et~al.(2025)Li, Liu, Ding, Fan, Li, and Li}]{li2025large}
Jiatong Li, Wei Liu, Zhihao Ding, Wenqi Fan, Yuqiang Li, and Qing Li. 2025.
\newblock Large language models are in-context molecule learners.
\newblock \emph{IEEE Transactions on Knowledge and Data Engineering}.

\bibitem[{Li et~al.(2024)Li, Liu, Fan, Wei, Liu, Tang, and
  Li}]{li2024empowering}
Jiatong Li, Yunqing Liu, Wenqi Fan, Xiao-Yong Wei, Hui Liu, Jiliang Tang, and
  Qing Li. 2024.
\newblock Empowering molecule discovery for molecule-caption translation with
  large language models: A chatgpt perspective.
\newblock \emph{IEEE transactions on knowledge and data engineering}.

\bibitem[{Liang et~al.(2023)Liang, Zhang, Zhang, and Xie}]{liang2023drugchat}
Youwei Liang, Ruiyi Zhang, Li~Zhang, and Pengtao Xie. 2023.
\newblock Drugchat: towards enabling chatgpt-like capabilities on drug molecule
  graphs.
\newblock \emph{arXiv preprint arXiv:2309.03907}.

\bibitem[{Lin(2004)}]{lin2004rouge}
Chin-Yew Lin. 2004.
\newblock Rouge: A package for automatic evaluation of summaries.
\newblock In \emph{Text summarization branches out}, pages 74--81.

\bibitem[{Liu et~al.(2024)Liu, Feng, Xue, Wang, Wu, Lu, Zhao, Deng, Zhang, Ruan
  et~al.}]{liu2024deepseek}
Aixin Liu, Bei Feng, Bing Xue, Bingxuan Wang, Bochao Wu, Chengda Lu, Chenggang
  Zhao, Chengqi Deng, Chenyu Zhang, Chong Ruan, et~al. 2024.
\newblock Deepseek-v3 technical report.
\newblock \emph{arXiv preprint arXiv:2412.19437}.

\bibitem[{Liu et~al.(2023{\natexlab{a}})Liu, Nie, Wang, Lu, Qiao, Liu, Tang,
  Xiao, and Anandkumar}]{liu2023multi}
Shengchao Liu, Weili Nie, Chengpeng Wang, Jiarui Lu, Zhuoran Qiao, Ling Liu,
  Jian Tang, Chaowei Xiao, and Animashree Anandkumar. 2023{\natexlab{a}}.
\newblock Multi-modal molecule structure--text model for text-based retrieval
  and editing.
\newblock \emph{Nature Machine Intelligence}, 5(12):1447--1457.

\bibitem[{Liu et~al.(2023{\natexlab{b}})Liu, Zhang, Xia, Wu, Xie, Qin, Zhang,
  and Liu}]{liu-etal-2023-molxpt}
Zequn Liu, Wei Zhang, Yingce Xia, Lijun Wu, Shufang Xie, Tao Qin, Ming Zhang,
  and Tie-Yan Liu. 2023{\natexlab{b}}.
\newblock \href {https://doi.org/10.18653/v1/2023.acl-short.138} {{M}ol{XPT}:
  Wrapping molecules with text for generative pre-training}.
\newblock In \emph{Proceedings of the 61st Annual Meeting of the Association
  for Computational Linguistics (Volume 2: Short Papers)}, pages 1606--1616,
  Toronto, Canada. Association for Computational Linguistics.

\bibitem[{Liu et~al.(2023{\natexlab{c}})Liu, Li, Luo, Fei, Cao, Kawaguchi,
  Wang, and Chua}]{liu2023molca}
Zhiyuan Liu, Sihang Li, Yanchen Luo, Hao Fei, Yixin Cao, Kenji Kawaguchi, Xiang
  Wang, and Tat-Seng Chua. 2023{\natexlab{c}}.
\newblock Molca: Molecular graph-language modeling with cross-modal projector
  and uni-modal adapter.
\newblock \emph{arXiv preprint arXiv:2310.12798}.

\bibitem[{Lu et~al.(2024)Lu, Cao, Liu, Bai, Chen, Yao, Zheng, and
  Li}]{lu2024moleculeqa}
Xingyu Lu, He~Cao, Zijing Liu, Shengyuan Bai, Leqing Chen, Yuan Yao, Hai-Tao
  Zheng, and Yu~Li. 2024.
\newblock Moleculeqa: A dataset to evaluate factual accuracy in molecular
  comprehension.
\newblock \emph{arXiv preprint arXiv:2403.08192}.

\bibitem[{Luo et~al.(2023)Luo, Yang, Hong, Liu, and Nie}]{luo2023molfm}
Yizhen Luo, Kai Yang, Massimo Hong, Xing~Yi Liu, and Zaiqing Nie. 2023.
\newblock Molfm: A multimodal molecular foundation model.
\newblock \emph{arXiv preprint arXiv:2307.09484}.

\bibitem[{Papineni et~al.(2002)Papineni, Roukos, Ward, and
  Zhu}]{papineni2002bleu}
Kishore Papineni, Salim Roukos, Todd Ward, and Wei-Jing Zhu. 2002.
\newblock Bleu: a method for automatic evaluation of machine translation.
\newblock In \emph{Proceedings of the 40th annual meeting of the Association
  for Computational Linguistics}, pages 311--318.

\bibitem[{Pei et~al.(2023)Pei, Zhang, Zhu, Wu, Gao, Wu, Xia, and
  Yan}]{pei2023biot5}
Qizhi Pei, Wei Zhang, Jinhua Zhu, Kehan Wu, Kaiyuan Gao, Lijun Wu, Yingce Xia,
  and Rui Yan. 2023.
\newblock Biot5: Enriching cross-modal integration in biology with chemical
  knowledge and natural language associations.
\newblock In \emph{Proceedings of the 2023 Conference on Empirical Methods in
  Natural Language Processing}, pages 1102--1123.

\bibitem[{Su et~al.(2022)Su, Du, Yang, Zhou, Li, Rao, Sun, Lu, and
  Wen}]{su2022molecular}
Bing Su, Dazhao Du, Zhao Yang, Yujie Zhou, Jiangmeng Li, Anyi Rao, Hao Sun,
  Zhiwu Lu, and Ji-Rong Wen. 2022.
\newblock A molecular multimodal foundation model associating molecule graphs
  with natural language.
\newblock \emph{arXiv preprint arXiv:2209.05481}.

\bibitem[{Team et~al.(2023)Team, Anil, Borgeaud, Alayrac, Yu, Soricut,
  Schalkwyk, Dai, Hauth, Millican et~al.}]{team2023gemini}
Gemini Team, Rohan Anil, Sebastian Borgeaud, Jean-Baptiste Alayrac, Jiahui Yu,
  Radu Soricut, Johan Schalkwyk, Andrew~M Dai, Anja Hauth, Katie Millican,
  et~al. 2023.
\newblock Gemini: a family of highly capable multimodal models.
\newblock \emph{arXiv preprint arXiv:2312.11805}.

\bibitem[{Weininger(1988)}]{weininger1988smiles}
David Weininger. 1988.
\newblock Smiles, a chemical language and information system. 1. introduction
  to methodology and encoding rules.
\newblock \emph{Journal of chemical information and computer sciences},
  28(1):31--36.

\bibitem[{White et~al.(2023)White, Hocky, Gandhi, Ansari, Cox, Wellawatte,
  Sasmal, Yang, Liu, Singh et~al.}]{white2023assessment}
Andrew~D White, Glen~M Hocky, Heta~A Gandhi, Mehrad Ansari, Sam Cox, Geemi~P
  Wellawatte, Subarna Sasmal, Ziyue Yang, Kangxin Liu, Yuvraj Singh, et~al.
  2023.
\newblock Assessment of chemistry knowledge in large language models that
  generate code.
\newblock \emph{Digital Discovery}, 2(2):368--376.

\bibitem[{Ye et~al.(2025)Ye, Cai, Lai, Wang, Huang, Wang, Liu, and
  Zeng}]{ye2025drugassist}
Geyan Ye, Xibao Cai, Houtim Lai, Xing Wang, Junhong Huang, Longyue Wang, Wei
  Liu, and Xiangxiang Zeng. 2025.
\newblock Drugassist: A large language model for molecule optimization.
\newblock \emph{Briefings in Bioinformatics}, 26(1):bbae693.

\bibitem[{Yik and Dood(2024)}]{yik2024chatgpt}
Brandon~J Yik and Amber~J Dood. 2024.
\newblock Chatgpt convincingly explains organic chemistry reaction mechanisms
  slightly inaccurately with high levels of explanation sophistication.
\newblock \emph{Journal of Chemical Education}, 101(5):1836--1846.

\bibitem[{Yu et~al.(2024)Yu, Baker, Chen, Ning, and Sun}]{yu2024llasmol}
Botao Yu, Frazier~N Baker, Ziqi Chen, Xia Ning, and Huan Sun. 2024.
\newblock Llasmol: Advancing large language models for chemistry with a
  large-scale, comprehensive, high-quality instruction tuning dataset.
\newblock \emph{arXiv preprint arXiv:2402.09391}.

\bibitem[{Zhang et~al.(2024)Zhang, Liu, Tan, Chen, Yan, Yan, Li, Huang, Yue,
  Ouyang et~al.}]{zhang2024chemllm}
Di~Zhang, Wei Liu, Qian Tan, Jingdan Chen, Hang Yan, Yuliang Yan, Jiatong Li,
  Weiran Huang, Xiangyu Yue, Wanli Ouyang, et~al. 2024.
\newblock Chemllm: A chemical large language model.
\newblock \emph{arXiv preprint arXiv:2402.06852}.

\end{thebibliography}
\onecolumn
\appendix
\setlength{\parskip}{3pt}
\setlength{\itemsep}{0pt}
\section*{Appendix}
\section{Reproducibility}
The codes are available at \url{https://github.com/HeinzVonHank/MolErr2Fix}.\\
The dataset is available at \url{https://huggingface.co/datasets/YoungerWu/MolErr2Fix}.

\section{Standard Annotation Workflow}
The Standard Annotation Workflow presented establishes a systematic approach for error identification, classification, correction, and validation in chemical structure annotation. This workflow ensures dataset consistency and quality by implementing a hierarchical decision tree that evaluates chemical structure inputs through multiple verification paths. The methodology enables robust error detection and classification, ultimately enhancing the reliability of chemical structure datasets for computational chemistry applications.

\begin{figure}[htbp]
  \centering
  \includegraphics[width=1\textwidth]{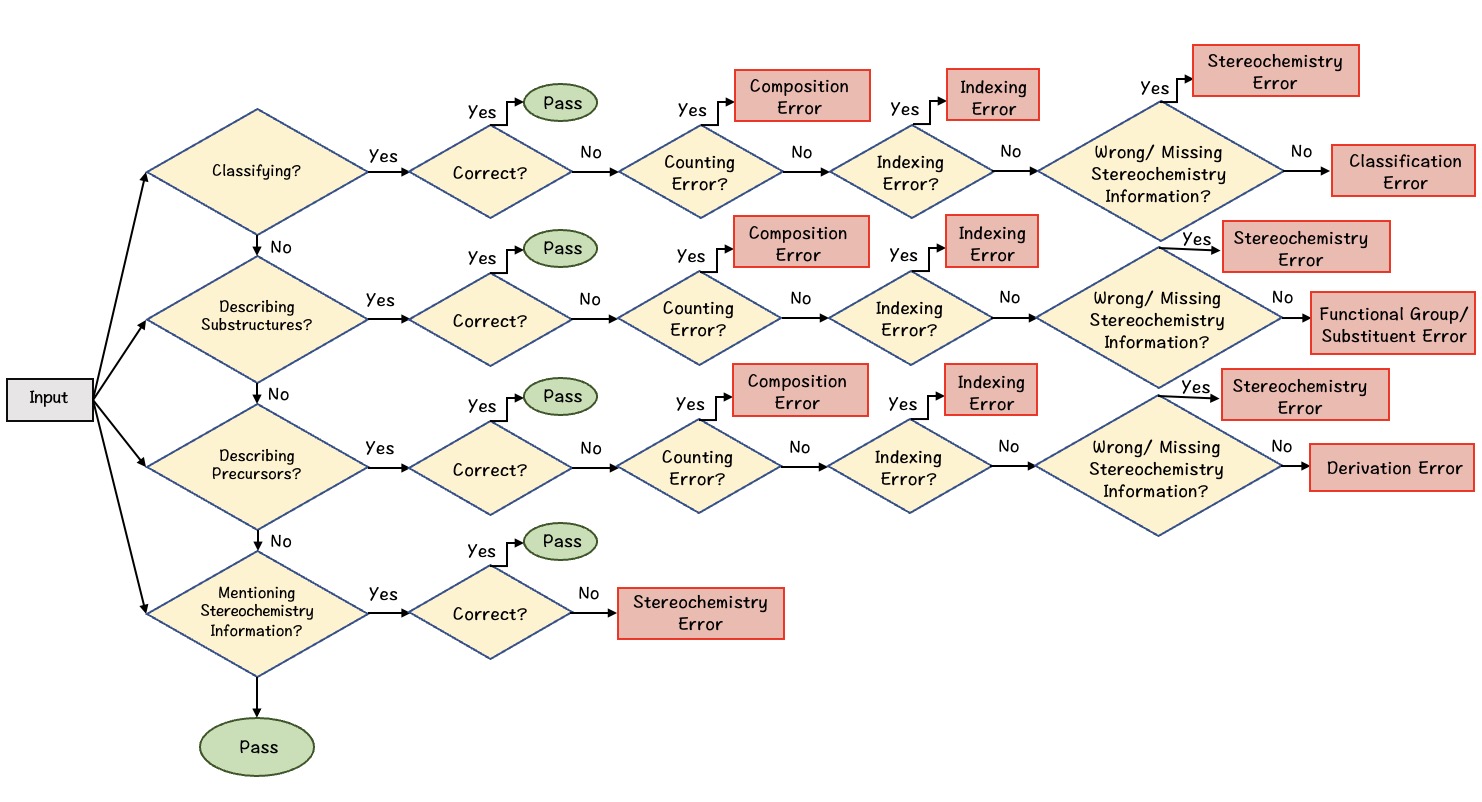}
  \caption{Standard annotation workflow illustrating the steps for error identification, classification, correction, and validation to ensure the dataset's consistency and quality.}
  \label{fig:anno}
\end{figure}

\subsection{Initial Input Classification}
Input chemical structures enter the workflow and are immediately sorted into one of four primary classification categories: classifying structures, describing substructures, describing precursors, or mentioning stereochemistry information. This initial categorization determines the subsequent verification path for each input, allowing for targeted error analysis based on the specific context of the chemical structure representation.\\

\subsection{Verification Paths}

\paragraph{Path 1: Classifying Structures}
When an input is identified as classifying a chemical structure, it first undergoes a correctness check that evaluates the structure for overall accuracy. If correct, the structure receives a pass designation. If incorrect, the annotation proceeds to a detailed error analysis sequence. The error analysis begins with checking for counting errors in atomic counts or stoichiometry, which would result in a composition error classification. If no counting errors are found, the process continues to examine for indexing errors in atomic or group indexing, which would lead to an indexing error designation. In the absence of indexing errors, the system checks for problems with stereochemistry information. If stereochemical information is wrong or missing, a stereochemistry error is identified; otherwise, the issue is classified as a general classification error.

\paragraph{Path 2: Describing Substructures}
For inputs categorized as describing chemical substructures, the workflow first verifies the overall correctness of the substructure description. Accurate descriptions pass immediately, while inaccurate ones undergo thorough error analysis. The system first examines potential counting errors in component counts, which would be classified as composition errors. If component counts are accurate, the workflow proceeds to check for positional indexing errors. When found, these are labeled as indexing errors. If the indexing is correct, the system evaluates the stereochemistry information for accuracy and completeness. Errors in stereochemistry are appropriately labeled, while issues without stereochemical involvement are classified as functional group or substituent errors.

\paragraph{Path 3: Describing Precursors}
When evaluating chemical precursor descriptions, the workflow first assesses overall accuracy. Correct descriptions pass validation immediately. Incorrect descriptions undergo a sequential error analysis beginning with precursor component counts, where inaccuracies result in composition errors. With accurate component counts, the system examines precursor indexing, classifying any issues as indexing errors. If indexing is correct, the workflow evaluates stereochemistry information, with inaccuracies or omissions classified as stereochemistry errors. Precursor descriptions with accurate stereochemistry information but other inaccuracies are classified as derivation errors, reflecting issues with the synthetic pathway representation.

\paragraph{Path 4: Mentioning Stereochemistry Information}
The final verification path addresses inputs specifically focused on stereochemistry information. The workflow simply evaluates the accuracy of the provided stereochemistry information. Correct stereochemical descriptions pass validation, while incorrect ones are classified as stereochemistry errors. Importantly, if no stereochemistry information is mentioned in an input that was expected to contain such information, the annotation receives a pass designation, as the absence of stereochemistry information is not considered an error in this context.\\

\subsection{Error Categories}

The workflow identifies and classifies errors into six comprehensive categories that encompass the range of possible annotation issues. Composition errors represent inaccuracies in the atomic or molecular composition, including incorrect counting of atoms, groups, or molecular components. Indexing errors encompass mistakes in the positional or sequential indexing of atoms, groups, or structural elements, which can significantly impact the interpretation of chemical structures. Stereochemistry errors indicate incorrect or missing information regarding the three-dimensional arrangement of atoms or groups in a molecule, including chirality, cis/trans isomerism, or other stereochemical properties critical for proper structural representation. Classification errors represent fundamental misconceptions in the identification of the chemical structure. Functional group or substituent errors involve mistakes in the identification, positioning, or description of functional groups or substituents that define chemical reactivity. Finally, derivation errors relate to inaccuracies in the representation of synthetic pathways of described precursors.\\

\subsection{Implementation and Quality Assurance}

The implementation of our standard annotation workflow follows a comprehensive approach to ensure data quality. Each chemical structure annotation undergoes preliminary analysis to determine its primary category, establishing the appropriate verification path. The annotation then proceeds through sequential verification steps based on its categorization, with potential errors identified and classified according to the defined error categories. When errors are detected, specific correction protocols are triggered based on the error type, ensuring appropriate remediation strategies. Corrected annotations undergo revalidation through the workflow to confirm all errors have been addressed, ensuring the integrity of the correction process. Only annotations that successfully pass all verification checks receive final approval for inclusion in the dataset, maintaining the highest standards of data quality.\\

\section{Error Typologies in Molecular Descriptions}

\subsection{Functional Group/Substituent Error}
\textbf{Definition:} The LLM incorrectly described a substructure within a given molecule, including misidentifying the type or name of substituents or functional groups, incorrectly describing the connectivity between substructures, or mistakenly claiming the presence of a substructure that does not actually exist. These errors are most commonly found in descriptions that begin with phrases such as 'The molecule has...' or 'The molecule contains...'

\textbf{Example 1:} \\
\textit{Input SMILES:} C1=CC(=CC=C1C[C@@H](C(=O)O)NS(=O)(=O)O)O \\
\textit{Problematic Description:} The molecule contains a carboxylic acid group, a sulfonamide group, and a phenol group. \\
\textit{Problem:} The substructure attached to the beta carbon of the carboxyl group in this molecule does not satisfy the definition of a sulfonamide group; therefore, we cannot say that the molecule contains a sulfonamide group. Instead, since the substructure connected to the nitrogen atom satisfies the definition of a sulfo group, we can say that the molecule contains a sulfo group.

\textbf{Example 2:} \\
\textit{Input SMILES:} CC(=CCC/C(=C/CC/C(=C/CC[C@@]1(CCC2=C(O1)C=CC(=C2)O)C)/C)/C)C \\
\textit{Problematic Description:} The molecule is a natural product with a furan ring fused to a phenol moiety. \\
\textit{Problem:} First, the fused ring structure has the name chromane. Second, even if you describe it as fused rings, the ring with oxygen is pyranose instead of furan because it contain six members

\textbf{Example 3:} \\
\textit{Input SMILES:} CC(=CCC/C(=C/CC/C(=C/CC[C@@]1(CCC2=C(O1)C=CC(=C2)O)C)/C)/C)C \\
\textit{Problematic Description:} It contains multiple alkene groups in a conjugated system with several methyl substituents along the carbon chain. \\
\textit{Problem:} The multiple alkene groups are not in a conjugated system

\subsection{Classification Error}
\textbf{Definition:} The LLM incorrectly described the name, type, or category of the species, including incorrect statements about whether the molecule is aromatic. These errors are most commonly found in descriptions that begin with phrases such as 'The molecule is...' or 'The molecule belongs to...'

\textbf{Example 1:} \\
\textit{Input SMILES:} C1=CC(=C(C=C1O)[O-])C2=COC3=CC(=CC(=C3C2=O)O)O \\
\textit{Problematic Description:} The molecule is a flavonoid, specifically a flavonol. \\
\textit{Problem:} The input molecule should be classified as an isoflavone instead of flavonol
.

\textbf{Example 2:} \\
\textit{Input SMILES:} C1=COC=CO1 \\
\textit{Problematic Description:} The molecule is 1,4-dioxin (1,4-dioxacyclohexa-2,5-diene), an aromatic heterocycle. \\
\textit{Problem:} This statement incorrectly describes the molecule as aromatic. In reality, 1,4-dioxine is a non-aromatic oxacycle.

\textbf{Example 3:} \\
\textit{Input SMILES:} CCCCCC(/C=C/C=C\textbackslash C/C=C\textbackslash C/C=C\textbackslash CCCC(=O)O)O \\
\textit{Problematic Description:} This compound belongs to the class of omega-3 fatty acids. \\
\textit{Problem:} This compound does not belong to the class of omega-3 fatty acids.

\subsection{Derivation Error}
\textbf{Definition:} The LLM incorrectly described the precursor, particularly when it identified a precursor that does not share the core structure of the given molecule, therefore not qualifying as a true derivative, or when it stated an incorrect conjugate acid or base. These errors are most commonly found in descriptions such as 'The molecule is a XXX derivative...', 'The molecule is derived from...', or 'The molecule is a conjugate acid of...'

\textbf{Example 1:} \\
\textit{Input SMILES:} NP(O)O \\
\textit{Problematic Description:}  The simplest phosphoric monoamide derivative of phosphoric acid. \\
\textit{Problem:} This description incorrectly identifies the compound as a derivative of phosphoric acid (H$_3$PO$_4$), whereas the correct molecule is a derivative of phosphorous acid (H$_3$PO$_3$).

\textbf{Example 2:} \\
\textit{Input SMILES:} CCN \\
\textit{Problematic Description:} It is a conjugate base of a N-methylputrescinium(2+). \\
\textit{Problem:} The conjugate acid is misidentified. Ethanamine, when protonated, forms the ethylaminium ion (CH$_3$CH$_2$NH$_3^+$), not an N‑methylputrescinium ion, which would belong to a different chemical family.

\textbf{Example 3:} \\
\textit{Input SMILES:} C1=CC(=C(C=C1C2(C3=C(C(=C(C(=C3Br)Br)Br)Br)C(=O)O2)C4=CC(=C(C=C4)O)S (=O)(=O)[O-])S(=O)(=O)[O-])O.[Na+].[Na+] \\
\textit{Problematic Description:} It is a derivative of eosin Y, which is a fluorescent red dye commonly used as a biological stain. \\
\textit{Problem:}  It is not a derivative of eosin Y. Instead, it derives from a 2-benzofuran-1(3H)-one.

\subsection{Stereochemistry Error}
\textbf{Definition:} The LLM incorrectly identified the stereochemical configuration, such as making incorrect statements about the R/S, Z/E, or cis/trans configuration of a given molecule. In some cases, it failed to specify the stereochemical configuration even though it could be determined from the input SMILES representation of the molecule.

\textbf{Example 1:} \\
\textit{Input SMILES:} CCCCCCCCCCCCCCCC/C=C\textbackslash OC[C@H](COP(=O)(O)O)O \\
\textit{Problematic Description:} It consists of a glycerol backbone with a phosphate group at the sn-3 position, a hydroxyl group at the sn-2 position, and a trans-configured monounsaturated fatty acid (likely a C18:1 fatty acid) attached via an ether linkage at the sn-1 position. \\
\textit{Problem:} This molecule is a cis-configured monounsaturated fatty acid, not trans-configured.

\textbf{Example 2:} \\
\textit{Input SMILES:} CCOC1=C(C=CC(=C1[C@@H(CS(=O(=O)C)N2C(=O)C3=C(C2=O)C(=CC=C3)NC (=O)C)OC \\
\textit{Problematic Description:} Featuring a methoxy-substituted phenyl ring and a chiral center. \\
\textit{Problem:} Although the description notes the presence of a chiral center, it does not specify the stereochemistry. The correct structure includes a (1S)-configuration on the 1-(3-ethoxy-4-methoxyphenyl) portion, which is featuring a (1S)-configured 1-(3-ethoxy-4-methoxyphenyl) group.

\textbf{Example 3:} \\
\textit{Input SMILES:} C1CSS[C@H]1CCCCC(=O)O \\
\textit{Problematic Description:} The molecule is the (R)-enantiomer of lipoic acid. \\
\textit{Problem:} The description misassigns the stereochemistry. According to the correct structure, the molecule is the (S)-enantiomer of lipoic acid (i.e. 5-[(3S)-dithiolan-3-yl]pentanoic acid). This is critical because the (S)-enantiomer is the enantiomer of naturally occurring (R)-lipoic acid and may exhibit different biological effects.

\subsection{Sequence/Composition Error}
\textbf{Definition:} The LLM generated an incorrect name due to errors in counting the length of certain components within the given molecule, including incorrect statements about the number of substituents, the length of the main carbon chain, or the size of n-membered rings.

\textbf{Example 1:} \\
\textit{Input SMILES:} C1[C@@H]([C@@H]([C@H](N1)C(=O)O)CC(=O)O)C2=CC=C(NC2=O)C(=O)O \\
\textit{Problematic Description:} A monocarboxylic acid. \\
\textit{Problem:} The description indicates that the molecule has only one carboxyl group, whereas the correct structure is a tricarboxylic acid. In the true structure, the pyrrolidinecarboxylic acid moiety includes a carboxyl group on the pyrrolidine ring, plus an additional carboxyl function on the pyridone portion and a carboxymethyl substituent—totaling three carboxyl groups.

\textbf{Example 2:} \\
\textit{Input SMILES:} CCC(CC)O \\
\textit{Problematic Description:} The molecule is 2-butanol … with the chemical formula C$4$H${10}$O. \\
\textit{Problem:} The description misidentifies the molecule’s carbon chain length. 2-Butanol has a four‑carbon backbone, whereas the correct structure is pentan-3-ol, which has a five‑carbon chain with the hydroxyl group at the third carbon.

\textbf{Example 3:} \\
\textit{Input SMILES:} C1CCC(=NNC2=NC(=CS2)C3=CC=C(C=C3)Cl)C1 \\
\textit{Problematic Description:} The structure contains a cyclohexane ring, a hydrazone group, a thiazole heterocycle, and a chlorophenyl moiety. \\
\textit{Problem:} It consist of a cyclopentane ring instead of a cyclohexane ring.\\

\subsection{Indexing Error}
\textbf{Definition:} The LLM incorrectly described the position of a substituent within the given molecule, including incorrect use of conventional ring labeling systems—such as the A/B/C ring assignments in flavonoids—or misapplication of positional terms like ortho, meta, and para in disubstituted benzene rings.

\textbf{Example 1:} \\
\textit{Input SMILES:} CCCCCCCC[C@@H](/C=C/CCCCCCC(=O)O)OO \\
\textit{Problematic Description:} (9E,11E)-octadeca-9,11-dienoic acid \\
\textit{Problem:} This statement incorrectly specifies that the fatty acid has double bonds at positions 9 and 11 with E configurations. In contrast, the correct structure is derived from (8E)-octadecenoic acid, which contains double bond at position 8 in the E configuration.

\textbf{Example 2:} \\
\textit{Input SMILES:} C1=CC(=C(C=C1O)[O-])C2=COC3=CC(=CC(=C3C2=O)O)O \\
\textit{Problematic Description:} With hydroxyl groups at positions 3, 5, and 7 on the A and C rings, and a deprotonated hydroxyl group at position 4' on the B ring. \\
\textit{Problem:} By convention, the C ring of flavonoids refers to the central pyran ring, not the branched phenyl ring. Therefore, there is no hydroxyl group on the C ring of this molecule. Instead, the correct positions of the hydroxyl groups are at positions 5 and 7 on the A ring, and at position 4$^\prime$ on the B ring. The deprotonated hydroxyl group is located at position 2$^\prime$ on the B ring.

\textbf{Example 3:} \\
\textit{Input SMILES:} C1=CC(=CC=C1/C=C/C(=O)O)C(F)(F)F \\
\textit{Problematic Description:} The molecule is a member of the class of (trifluoromethyl)benzenes, consisting of trans-cinnamic acid having a trifluoromethyl substituent at the meta-position. \\
\textit{Problem:} The trifluoromethyl group (-CF3) is not at the meta-position. According to the correct structure, the trifluoromethyl group is at the para-position relative to the cinnamic acid group. It misidentifies the position of the trifluoromethyl substituent. The para-position is directly opposite the cinnamic acid group on the benzene ring.

\section{Model Prompts}
\subsection{Prompts for Caption generation by LLMs}
You are now working as an excellent expert in chemistry and drug discovery. Given the SMILES representation of a molecule, your job is to predict the caption of the molecule. The molecule caption is a sentence that describes the molecule, which mainly describes the molecule's structures, properties, and production.

Example:

Instruction: Given the SMILES representation of a molecule, predict the caption of the molecule.

Input: CCCCCCCCCCCCCCCC(=O)OC(CCCCCCCCC)CCCCCCCC(=O)O

Your output should be: \{"caption": "The molecule is a FAHFA (fatty acid ester of a hydroxy fatty acid) obtained by formal condensation of the carboxy group of palmitic acid with the hydroxy group of 9-hydroxyoctadecanoic acid (9-hydroxystearic acid). It has a role as a human metabolite, a hypoglycemic agent, and an anti-inflammatory agent. It is a FAHFA and a long-chain fatty acid. It derives from a hexadecanoic acid and an octadecanoic acid. It is a conjugate acid of a 9-PAHSA(1-)."\}

Your response should only be in the JSON format above; THERE SHOULD BE NO OTHER CONTENT INCLUDED IN YOUR RESPONSE.

\subsection{Prompts for Error Detection}
You are an expert in molecular chemistry and error detection. Your task is to analyze a given molecule 
and determine which error types are present.

List of Possible Error Types:
\$error\_types\_info

Molecule Information:

Molecule's structure (SMILES format):\$smiles
Molecule's Description: \$description

Task:
Based on the provided structure and description, carefully examine whether any of the error types listed above exist 
in this molecule.\\
Please return ONLY a list of error codenames (e.g., ["E1","E2"]). DO NOT add anything else.\\
Response Format:
Your response must be a valid string that can be converted to list by the eval function in Python. If no error is detected, return an empty list \texttt{[]}.

\subsection{Prompts for Error Localization}
You are an expert in molecular chemistry and error detection.\\
Below are the seven error types with their definitions:\\
\{error\_types\_info\}\\
\\
Given the following molecule information and its erroneous description, list all the errors present in the description.\\
For each error, output the error type and the exact text segment (error span) from the description.\\
Please output the result in JSON format as an array of objects, each object having the keys ``error\_type'' and ``error\_span''.\\
Do not include any additional commentary.\\
\\
Molecule's structure (SMILES): \{smiles\}\\
Erroneous Description: \{description\}

\subsection{Prompts for Error Explanation}
You are a chemistry expert specialized in molecular structure understanding and error detection.\\
\\
You are given a molecule and a faulty description of it. A specific fragment in the description is believed to be incorrect. Your task is to explain \textbf{why} that part is wrong based on the molecule's actual structure.\\
\\
Molecule Information:
SMILES: \{smiles\}

Faulty Description:\\
``\{description\}''\\
\\
Suspected Faulty Fragment:\\
``\{error\_span\}''\\
\\
Your Task:\\
Explain briefly and clearly why the above fragment is incorrect. Limit your explanation to 1-2 concise sentences.

\subsection{Prompts for Evaluation of Error Explanation}

You are a strict evaluator. Please read the two statements below, which describe 
the reason behind a specific chemical error. Determine if they convey the same meaning.\\
\\
If they do, respond with ``Yes'' (and nothing else).\\
If they do not, respond with ``No'' (and nothing else).\\
\\
Statement A: ``\{sentA\}''\\
{Statement B: ``\{sentB\}''

\subsection{Prompts for Error Revision}
You are an expert in molecular structure and error correction.\\
The molecule and its description below contain an error, and we have pointed out the erroneous segment for you.\\
\\
Molecule's structure (SMILES): \$smiles\\
Erroneous Description: \$description\\
Erroneous segment: ``\$wrong\_segment''\\
\\
Please correct the error by providing a corrected substitution for the pointed out error segment, 
your answer should have a similar length as the erroneous segment,
return the corrected segment only, and do not include any other text.

\subsection{Prompts for Evaluation of Error Revision}
You are an expert in molecular structure and error correction evaluation.\\
Below is the original erroneous description:
``\$description''\\
\\
An error has been identified in the description as: ``\$wrong\_segment''\\
The human-corrected version for this error is: ``\$human\_correct''\\
The model-generated correction for this error is: ``\$llm\_correct''\\
\\
Please evaluate whether the model-generated correction properly fixes the error by replacing the wrong segment with a fragment that conveys the same meaning as the human correction, and without modifying parts of the description that were not marked as errors.\\
If the correction is appropriate, reply with 1 (and nothing else). Otherwise, reply with 0 (and nothing else).

\section{Annotator Information}
Our annotation team consisted of two primary annotators—a 5th-year PhD and a 2nd-year Master's student, both from a university chemistry department. In addition, a professor in chemistry performed quality control and spot checks throughout the annotation process.


\end{document}